\documentclass[prd,preprint,amsmath,amssymb,longbibliography]{revtex4-1}
\usepackage[utf8]{inputenc}
\usepackage{xcolor}
\usepackage{graphicx}
\usepackage{natbib}
\usepackage{amsmath}
\usepackage{adjustbox}
\usepackage[graphicx]{realboxes}
\usepackage{setspace}
\usepackage[margin=1in]{geometry}
\usepackage{dcolumn}

\usepackage{lineno}
\usepackage{float}
\usepackage{tikz}
\usepackage{import}

\usepackage[colorinlistoftodos]{todonotes}

\usepackage[colorlinks=true,backref=false, linktocpage=true,
citecolor=blue,urlcolor=blue,linkcolor=blue,pdfpagemode=UseOutlines,unicode]{hyperref}

\hypersetup{%
  bookmarksnumbered=true,
  pdftitle = {},
  pdfsubject = {},
  pdfauthor = {},
  pdfkeywords = {}
}

\begin{document}

\title{\texorpdfstring{Self-consistent optimization of the $z$-Expansion for $B$ meson decays}{title}}

\author{Daniel Simons$^1$}
\author{Erik Gustafson$^2$}
\author{Yannick Meurice$^1$}
\affiliation{$^1$Department of Physics and Astronomy, The University of Iowa\\
Iowa City, IA 52242, USA}
\affiliation{$^2$Fermilab, Batavia, IL 60510, USA}
\date{\today}

\begin{abstract}
We discuss the self-consistency imposed by the analyticity of  regular parts of form factors, appearing in the $z$-expansion for semileptonic $B$-meson decays, when fitted in different kinematic regions. Relying on the uniqueness of functions defined by analytic continuation, 
we propose four metrics which measure the departure from the ideal analytic self-consistency. We illustrate the process using Belle data for $B\rightarrow D\ell \nu_\ell$. For this specific example, the metrics provide consistent indications that some choices (order of truncation, BGL or BCL) made in the form of the $z$-expansion can be optimized. However, other choices ($z$-origin, location of isolated poles and threshold constraints) appear to have very little effect on these metrics. 
We briefly discuss the implication for optimization of the $z$-expansion for nucleon form factors relevant for neutrino oscillation experiments.

\end{abstract}
\def\art{article}
\def\low{\mathcal L}
\def\high{\mathcal H}
\maketitle

\section{Introduction}


Experimental differential decay rates for exclusive  semileptonic decays of $B$-mesons \cite{BaBar:2007cke,BaBar:2007ddh,BaBar:2010efp,BaBar:2012thb,Belle:2010hep,
mainb2d,Belle:2018ezy,Belle-II:2020dyp,Belle-II:2020ewe},
combined with ab-initio lattice QCD calculations of the hadronic form factors \cite{Bailey:2008wp,Bernard:2009ke,flynn2023exclusive,FermilabLattice:2015mwy,FermilabLattice:2015cdh,Na_2015,Bailey:2015dka,MILC:2015uhg,FermilabLattice:2019qpy,FermilabLattice:2021cdg,Parrott:2020vbe,Parrott:2022rgu,Cooper:2021bkt,flynn2023exclusive,flynn2023bayesian}
provide reliable numerical estimations of the CKM matrix elements $|V_{ub}|$ and $|V_{cb}|$. 
Accurate lattice calculations are only possible for a large enough  invariant square of the 
4-momentum of the leptons, denoted $q^2$, or more specifically when the recoil energy of the final state meson is significantly smaller than the inverse lattice spacing. In order to predict the shape of the differential decay rate over the entire kinematic range from reliable lattice results in the high $q^2$ region, an analytic continuation method developed in the context of Kaon decays \cite{PhysRevD.4.725,PhysRevD.3.2807} has been adapted for $B$-meson decays by Boyd, Grinstein and Lebed (BGL)~\cite{originalBGL} and Bourrely, Caprini and Lellouch (BCL)~\cite{originalBCL}. The method is often called the $z$-expansion. 
The basic idea 
is to map the branch cut in the complex $q^2$ plane onto the boundary of the unit disk in $z$ with the rest of the cut complex $q^2$ plane being mapped into the interior of the disk. The goal is to find parameterizations of the form factors for specific processes where the effects of thresholds and isolated poles can to some extent be separated from a smooth behavior in the kinematic range. Ideally, after the mapping, the kinematic range  becomes a small interval near the origin and a few terms in the Taylor expansion provide reliable results. 
General strategies for combining the lattice and experimental data are discussed in Ref. \cite{Bernard:2009ke}. 

The extrapolation of lattice results with computationally accessible $q^2$ to the full kinematic range relevant for experimental analysis has been performed for various decay modes and by various collaborations 
\cite{Bailey:2008wp,Bernard:2009ke,flynn2023exclusive,FermilabLattice:2015mwy,FermilabLattice:2015cdh,Na_2015,MILC:2015uhg,Bailey:2015dka,FermilabLattice:2019qpy,FermilabLattice:2021cdg,Parrott:2020vbe,Parrott:2022rgu,Cooper:2021bkt,flynn2023exclusive,flynn2023bayesian}. 
Specific choices will be reviewed below. 
In general, the agreement with the overall shape of the experimental differential decay rate provides a strong guidance to select reasonable procedures. If one assumes the standard model is correct then an ab-initio calculation in the full kinematic range should reproduce the shape of the experimental data. Under this assumption, the only unknown quantity is $V_{cb}$, a Cabibbo-Kobayashi-Maskawa matrix element.
The $z$-expansion being a compact and model-independent method is very important to summarize the experimental results, especially as it does not depend on the binning procedure.
Recent experiments provide fits of their data using the $z$-expansion. This amounts to continuous functions that allow comparisons among experiments with different binnings.
For semileptonic decays of $B$-mesons involving tree-level virtual $W^{\pm}$ bosons, the form factor can  be expressed in term of {\it analytic} functions in the entire kinematic interval. An important implication is that a (perfect) knowledge of the analytic function in any open set in the complex $q^2$ plane {\it uniquely} determines the function in the whole interval provided that no singularities or cuts prevent the analytical continuation~\cite{knopp2013theory}.

In this \art, we discuss the self-consistency imposed by analyticity on regular parts of form factors when fitted in different kinematic regions. In Sec. \ref{sec:theory}, we review the BGL and BCL parmeterizations. In Sec. \ref{sec:results}, we consider existing goodness of fit measures ($\chi^2$ and AIC) and 
define four dimensionless metrics which measure the departure from ideal analytic self-consistency. These metrics are ``cost functions" for which a large value indicate an inconsistent parameterization  conflicting with the assumed analyticity as defined mathematically  in \cite{knopp2013theory}.
We illustrate the idea by calculating these four metrics for $B\rightarrow D\ell \nu_\ell$ using partial decays widths provided by the Belle collaboration \cite{mainb2d}. 
The numerical results are analyzed in Sec. \ref{sec:analysis} where we discuss the possibility of discriminating among a certain number of choices (order of truncation, BGL or BCL, $z$-origin and threshold constraints) made in the $z$-expansion.
The results are summarized in the conclusions where we also comment on
new methods of determining the order of truncation of the $z$-expansion \cite{Di_Carlo_2021,flynn2023bayesian}.
We also briefly possible applications for optimization of lattice nucleon form factors reviewed in \cite{Meyer:2022mix} and relevant for neutrino experiments such as DUNE.

\section{BGL and BCL parameterizations}
\label{sec:theory}

In the following, we focus on different parameterizations of the form factor that describe the decays $B^0 \rightarrow D^- \ell^+ \nu_{\ell}$  and $B^+ \rightarrow D^0 \ell^+ \nu_{\ell}$, with $\ell = \{e, \mu \}$. In the isospin limit, these processes can be described by a differential decay rate that depends on the hadronic recoil variable ${w\equiv (m_B^2+m_D^2-q^2)/(2m_B m_D)}$~\cite{mainb2d},
\begin{equation}
    \frac{d \Gamma}{dw} = K(w^2 - 1)^{3/2}  f_+(w)^2,
    \label{eqn:diff_decay}
\end{equation}
with
\begin{equation}
    K = \frac{G^2_F m^3_D}{48 \pi^3} |V_{cb}|^2(m_B + m_D)^2 \frac{4r}{(1+r)^2}  \eta^2_{EW} ,
    \label{eqn:diff_decay_constant}
\end{equation}
Where $G_F$ is the Fermi coupling constant, $m_B$ and $m_D$ are the masses of the $B$ and $D$ mesons respectively, $r = m_D / m_B$, and $\eta_{EW}$ represents the electroweak corrections.

The two parameterizations of the vector form factor $f_+(w)$ that we investigate are the BGL and BCL parameterizations. Both parameterizations use the $z$-expansion which takes the real kinematic range and embeds it into a complex domain, where the process of analytical continuation defines a unique regular (analytic and single-valued) function. The mapping variable is $z(q^2, t_0)$, where: 

\begin{equation}
    z(q^2, t_0) = \frac{\sqrt{t_+ - q^2}-\sqrt{t_+ - t_0}}{\sqrt{t_+ - q^2}+\sqrt{t_+ - t_0}}, 
\label{eq:zexp_equation}
\end{equation}

\noindent
$q^2$ is the momentum transfer, ${q^2 = m_B^2 + m_D^2 - 2 w m_B m_D}$, and $t_+ = (m_B + m_D)^2$. 
This change of coordinates maps the cut complex $q^2$ plane onto the unit disk. At threshold, $q^2=t_+$ and $z=-1$. The cut is mapped into the boundary of the disk. The variable $t_0$ determines where the $z$-expansion is centered about. We consider two $t_0$ values ${t_0 = t_{opt} = (m_B+m_D)(\sqrt{m_B} - \sqrt{m_D})^2}$ and ${t_0 = t_{-} = (m_B - m_D)^2}$. The choice of $t_0$ should not appreciably affect the $z$-expansion fit results, but it can be used to adjust the systematic uncertainties. Following the original authors, $t_0 = t_{opt}$ is used with BCL~\cite{originalBCL} and $t_0 = t_-$ is used with BGL~\cite{originalBGL}. The choice $t_0 = t_{opt}$ puts $z$ in the range $z \in [-0.0323, 0.0323]$, and the choice $t_0 = t_{-}$ puts z in the range $z \in [0.0, 0.0646]$.

We define the BGL parameterization as $f_{+,BGL}$, with the explicit form 
used for a lattice calculation \cite{MILC:2015uhg} and the analysis of the Belle data \cite{mainb2d}, both for  $B\rightarrow D\ell\nu_{\ell}$.
\begin{equation}
    f_{+,BGL}(z) \equiv \frac{1}{\phi_+(z)} \sum^N_{n=0} a_{+,n} z^n, \\
    \label{eq:fplus_BGL}
\end{equation}
\noindent
with 
\begin{eqnarray}
\phi_+(z) &=& 1.1213 (1+z)^2 (1-z)^{1/2} [(1+r)(1-z) + 2 \sqrt{r} (1+z)]^{-5}.
\end{eqnarray}
\noindent
The outer function, $\phi_{+}(z)$, is to some extent arbitrary but must be analytic and non-zero for $|z|<1$ in order to enforce the unitarity condition on $a_{+,n}$~\cite{originalBGL,mainb2d}. We then define the BCL parameterization as $f_{+,BCL}$, with the explicit form 
used for a lattice calculation \cite{FermilabLattice:2015mwy} $B\rightarrow \pi\ell\nu_{\ell}$ but with $m_{B^*}^2$ replaced by $m_{B_c^*}^2$.  

\begin{equation}
    f_{+,BCL}(z) \equiv \frac{1}{1 - q^2(z)/m_{B_c^*}^2} \sum^{K-1}_{k=0} b_{+,k} [z^k - (-1)^{k-K} \frac{k}{K} z^{K}]. \\
    \label{eq:fplus_BCL}
\end{equation}

One difference between the BGL and BCL parameterizations is the use of the threshold condition discussed in Appendix ~\ref{appendixC}. Here, BGL does not use the threshold condition while BCL does use it. The free parameters $a_{+,n}$ and $b_{+,n}$ are fitted by using least square fitting methods~\cite{lsqfit} and must satisfy the following unitarity conditions \cite{originalBGL}
\begin{equation}
    \sum^N_{n=0} |a_{+,n}|^2 \leq 1 ,\\
\end{equation}
\noindent
and \cite{originalBCL,FermilabLattice:2015mwy}
\begin{equation}
    \sum^{K}_{j,k=0} B_{jk} b_{+,j} b_{+,k} \leq 1 .\\
\end{equation}
\noindent
For $f_{+,BGL}$ we consider $N=0,1,2$ and for $f_{+,BCL}$ we consider $K=1,2,3$. It is important to note that $N$ is not the number of parameters while $K$ is, so to avoid confusion, the number of parameters used in the fit will be denoted $n_p$ for both parameterizations, where $n_p = N + 1$ and $n_p = K$. The $z^K$ term that is attached to every $b_{+,k}$ comes from the threshold condition which will be discussed in more detail in Appendix ~\ref{appendixC}.

The BGL parameterization sometimes includes a Blaschke factor $P_+(z)$ as well, which contains the information about the pole at $q^2_* \equiv m_{B_c^*}^2 = 40.02$ GeV$^2$. However, it has been shown that the Blaschke factor does not appreciably affect the $z$-fit for the BGL analysis of $B\rightarrow D\ell\nu_{\ell}$ due to the pole being very far away from the kinematical region~\cite{MILC:2015uhg}. For this reason the Blaschke factor has been set to 1 in \cite{MILC:2015uhg,mainb2d} and our definition in Eq. (\ref{eq:fplus_BGL}) follows this choice. While BCL replaces the commonly used outer function and Blaschke factor with a prefactor that has a pole at the same location as the Blaschke factor.  The pole $q^2_*$ corresponds to $z(q^2_{*},t_-) = -0.308$ and $z(q^2_{*},t_{opt}) = -0.337$. The construction of the $B_{jk}$ matrix can be found in~\cite{originalBCL}, and we calculate the values $B_{00}$, $B_{01}$, $B_{02}$ and $B_{03}$ for $B \rightarrow D \ell \nu_\ell$ and display them in Table~\ref{tab:Bmn_values}. The remaining $B_{mn}$ values can be calculated using the following relations~\cite{originalBCL},

\begin{equation}
    B_{j(j+k)} = B_{0k} ,\\
\end{equation}

\noindent
and

\begin{equation}
    B_{jk}=B_{kj} .\\
\end{equation}

\begin{table}[hbt!]
\caption{\label{tab:Bmn_values}
The matrix elements $B_{jk}$ which are used in the BCL unitarity condition for $n_p = 1,2,3$.}
\begin{ruledtabular}
\begin{tabular}{llll}
$B_{00}$ & $B_{01}$ & $B_{02}$ & $B_{03}$ \\
\colrule
0.0118 & -0.0028 & -0.0069 & 0.0038  \\
\end{tabular}
\end{ruledtabular}
\end{table}

The outer function and Blaschke factor are to some extent arbitrary so long that they are analytic and non-zero in the $z$ range that we are interested in. Following~\cite{MILC:2015uhg} where they set the Blaschke factor equal to one, it is then interesting to investigate the parameterizations of Eqs.~\eqref{eq:fplus_BGL} and~\eqref{eq:fplus_BCL} with their prefactors set equal to one. To differentiate the form factors when there are no prefactors, we denote Eq.~\eqref{eq:fplus_BGL} with no prefactors as $f_{+,\textrm{NN}}$ and we denote Eq.~\eqref{eq:fplus_BCL} with no prefactors as $f_{+,\textrm{NT}}$, where $\textrm{NN}$ stands for no-prefactor and no-threshold while $\textrm{NT}$ stands for no-prefactor with-threshold. The explicit forms of $f_{+,\textrm{NN}}$ and $f_{+,\textrm{NT}}$ can be found below,

\begin{equation}
    f_{+,\textrm{NN}}(z) = \sum^N_{n=0} a_{+,n} z^n, \\
\end{equation}

\noindent
and

\begin{equation}
    f_{+,\textrm{NT}}(z) = \sum^{N-1}_{n=0} b_{+,n} [z^n - (-1)^{n-N} \frac{n}{N} z^{N}] . \\
\end{equation}

\noindent
Where NN uses $t_0 = t_-$ and NT uses $t_0 = t_{opt}$.

 \def\df{\Delta f (z)}
 
\section{New metrics and results}
\label{sec:results}

We consider several tests that compare the goodness of fit and self-consistency of our models on the Belle data to determine the best parameterization with the best choices of input parameters. 

\subsection{\texorpdfstring{$\chi^2$ Test}{chi squared test}}

We use the LsqFit python library~\cite{lsqfit} to perform the fits of the different models with the Belle data, and the resulting fit parameters can be found in~\ref{appendixA}. The LsqFit library also provides the $\chi^2$ and reduced-$\chi^2$, $\chi^2_\nu$, and these values are provided for BGL, BCL, NN and NT with 1p, 2p and 3p in Table~\ref{tab:DiffDec_chi2}.

\begin{table}[hbt!]
\caption{\label{tab:DiffDec_chi2}
The $\chi^2$ and $\chi^2_{\nu}$ values calculated from the fits performed on the differential decay width data with the BGL, BCL, NN and NT parameterizations with 1p, 2p and 3p.}
\begin{ruledtabular}
\begin{tabular}{lll}
 & $\chi^2$ & $\chi^2_{\nu}$ \\
\colrule
BGL 1p & 99 & 11 \\ 
BGL 2p & 4.56 & 0.57 \\ 
BGL 3p & 4.55 & 0.65 \\
\hline
BCL 1p & 33.3 & 3.7 \\ 
BCL 2p & 4.64 & 0.58 \\ 
BCL 3p & 4.55 & 0.65 \\ 
\hline
NN 1p & 135 & 15 \\ 
NN 2p & 4.88 & 0.61 \\ 
NN 3p & 4.55 & 0.65 \\ 
\hline
NT 1p & 135 & 15 \\ 
NT 2p & 5.04 & 0.63 \\ 
NT 3p & 4.55 & 0.65 \\ 
\end{tabular}
\end{ruledtabular}
\end{table}

It is clear that the 2p and 3p cases provide significantly better models than the 1p case due to the extremely high $\chi^2$ and $\chi^2_{\nu}$ values in the 1p case. However it is not clear that the $\chi^2$ provides significant discrimination between 2p and 3p for a given model or discrimination among the different models. Similar remarks apply to $\chi^2_\nu$ which increases by about $10 \%$ when going from 2p to 3p, which could be due to the data having ten bins. This leads us to needing another metric to be able to discriminate between the models.

\subsection{Aikaike Information Criterion} 

A test that enables quantitative comparisons between models with differing numbers of parameters that aren't rigorously possibly without Bayesian techniques is the Aikaike Information Criterion (AIC). The AIC value is defined like an augmented $\chi^2$ value, where the augment is adding a $2 n_p$ term~\cite{AICdefinition_2020,AICdefinition_2022},

\begin{equation}
    \textrm{AIC} = 2 n_p + \chi^2 .\\
    \label{eq:AIC_final}
\end{equation}

\noindent
The inclusion of a penalty which is linear in $n_p$ is used to discourage overfitting, and the factor $2$ in front of $n_p$ is discussed in~\cite{AIC_original}. Therefore, the preferred model will be the model that has the lowest AIC value. Changes in the AIC values $\Delta \textrm{AIC}$ when the number of degrees of freedom $\nu$ is changed by $\Delta \nu$ can be considered significant if $|\Delta \textrm{AIC}| / \textrm{AIC}| > |\Delta \nu |/ \nu$~\cite{NPLQCD:2020ozd}.

These AIC values are displayed in Table~\ref{tab:DiffDec_AIC}. The $2 n_p$ term for 1p will not compensate for the drastically larger $\chi^2$ value compared to the 2p and 3p fits, clearly showing that 1p is not descriptive enough. The AIC values in Table~\ref{tab:DiffDec_AIC} show that all the 2p cases satisfy the inequality regarding $\nu$, which indicates that the 2p case is preferred over the 3p case. However, the AIC values between BGL and BCL for the same number of parameters is still too close to determine anything significant about which of the models does a better job of fitting the Belle data. We proceed to define our own metrics to find one that is able to distinguish between the different parameterization options that we consider. 

\begin{table}[hbt!]
\caption{\label{tab:DiffDec_AIC}
The AIC values calculated from the fits performed on the differential decay width data with the BGL, BCL, NN and NT parameterizations with 1p, 2p and 3p.}
\begin{ruledtabular}
\begin{tabular}{ll}
 & \multicolumn{1}{l}{AIC} \\
\colrule
BGL 1p & 101 \\ 
BGL 2p & 8.56 \\ 
BGL 3p & 10.55 \\
\hline
BCL 1p & 35.3 \\ 
BCL 2p & 8.64 \\ 
BCL 3p & 10.55 \\ 
\hline
NN 1p & 137 \\ 
NN 2p & 8.88 \\ 
NN 3p & 10.55 \\ 
\hline
NT 1p & 137 \\ 
NT 2p & 9.04 \\ 
NT 3p & 10.55 \\ 
\end{tabular}
\end{ruledtabular}
\end{table}

\subsection{Self-consistency metrics}

In the form factor expressions, the polynomials in $z$ are approximations of analytic functions in the kinematic range. The absence of singularities or cuts in that range implies that the exact knowledge of the function in an open region can uniquely determine the function in another region. This can be achieved by analytic continuation~\cite{knopp2013theory}. In the current context, if an analytic function is defined on an open segment of the  real $z$ axis corresponding to the kinematic range and if we partition this segment in a region $\high$ corresponding to a high-$z$ (or equivalently high-$w$ or low-$q^2$) part and the complementary region $\low$ in the low-$z$ region. It is then clear that ideally the perfect knowledge of the function in $\high$, uniquely determines the function in $\low$ and vice-versa.
 
In practice, if we use experimental data, we know the function at a finite number of points with a limited accuracy. It is expected that if we obtain a polynomial approximation in $\high$ using the data in  $\high$, that we call $f_+^{high}(z)$ and extend this polynomial to $\low$, and if we obtain $f_+^{low}(z)$ by swapping the roles of $\high$ and $\low$, then the discrepancy 
\begin{equation}
    \df\equiv f_+^{high}(z)-f_+^{low}(z),
\end{equation}
is nonzero and provide a measure of the inconsistency of the continuations due to imperfect knowledge of the function in addition to the uncertainty in the data.

A rough global measure of the inconsistency of a specific method used to obtain the polynomial approximation could be the $L^2$-norm of $\df$. This quantity depends on the 
units of the form factor and the range of $z$ in the integral. For a decent approximation, one would expect that $(\df)^2$ would be of the order of the average experimental variance $\bar{\sigma}^2_{exp}$ and we could expect to get a quantity of order one 
by dividing by the length of the $z$-interval and the average experimental variance $\bar{\sigma}^2_{exp} = 0.00199$. For these reasons we start with the dimensionless quantity
\begin{equation}
    C_0 \equiv \frac{1}{\bar{\sigma}^2_{exp} |z_{max}-z_{min}|} \int^{z_{max}}_{z_{min}} (\df)^2 dz, \\
\end{equation}

A more refined metric denoted  $C_1$, can be obtained by weighting \textit{locally} with the inverse local variance $\sigma^2_{exp}(z)$ obtained from the experimental data by interpolating with LsqFit. 

\begin{equation}
   C_1 \equiv \frac{1}{|z_{max}-z_{min}|} \int^{z_{max}}_{z_{min}} \frac{(\df)^2}{\sigma^2_{exp}(z)} dz \\
\end{equation}

\def\nb{n_{bin}}

When the experimental form factors are provided as binned data with $n_b$ bins, we can define a discrete version of $C_1$ as 

\begin{equation}
    D_1 = \frac{1}{\nb} \sum^{\nb}_{i=1} (\frac{\Delta f_i}{\sigma_i})^2, \\
\end{equation}

\noindent
with $\Delta f _i =\Delta f(z_i)$, $z_i$ being in the middle of the $i$-th bin. 
This can be calculated in a straightforward way without the need of interpolations. 
If the bins are narrow enough, we expect 
that $D_1\simeq C_1$.
The general form of $D_1$ is reminiscent of a $\chi$-square, however $\sigma_i^2$ is not the variance of $\Delta f_i$.

Given that the experimental binned data may involve significant correlations among the bins, we can pursue the analogy and generalize $D_1$ to 

\begin{equation}
   D_2 = \frac{1}{\nb} \sum^{\nb}_{i,j=1} \Delta f_i \mathcal{C}^{-1}_{ij} \Delta f_j,\\
\end{equation}

\noindent
with $\mathcal{C}_{ij}$ the covariance matrix of the binned data for the form factor.

The covariance matrix $\mathcal{C}_{ij}$ is calculated using sampled bootstrap form factor data points. Using the gvar python library~\cite{GvarLibrary}, we generated $M=10^4$ bootstrap differential decay width data sets generated from the Belle data using the underlying covariance matrix. Then using Eqs.~\eqref{eqn:diff_decay} and~\eqref{eqn:diff_decay_constant}, we converted the generated differential decay width data into data describing the form factor, with $f^i$ being the set of random form factor data in the $i$-th bin. Finally, we calculated $\mathcal{C}_{ij}$ using, 

\begin{equation}
    \mathcal{C}_{ij} = \frac{1}{M} \sum^{M}_{m=1} (f^i_m - \bar{f^i}) (f^j_m - \bar{f^j}) \\
    \label{eq:Cij_covariance}
\end{equation}

\noindent
Where $f^i_m$ is the m-th data point in the i-th bin and $\bar{f^i}$ is the mean of the data in the i-th bin. And the $\sigma^2_i$ from Eq.~\eqref{eq:Cij_covariance} is the diagonal entries of the covariance matrix, $\sigma^2_i = \mathcal{C}_{ii}$.

\begin{figure}
\begin{center}
        \includegraphics[width=\linewidth]{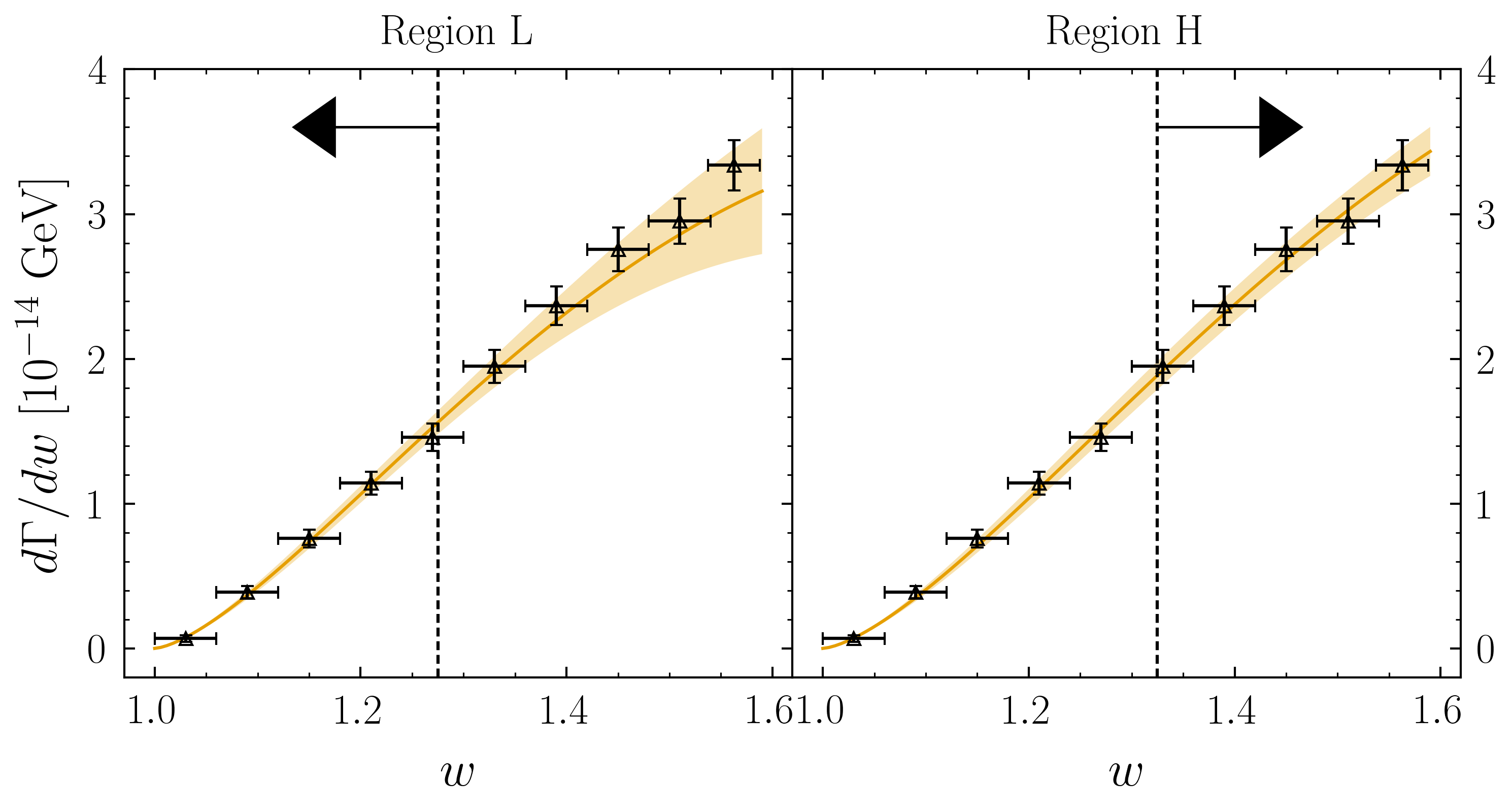}
\end{center}
        \caption{The results of our fits using the BGL parameterization. The red line is the mean value of our fit and the lighter red region is the 1-$\sigma$ error band, the black triangles are the Belle data with error bars, and the dashed arrow indicates whether region $\low$ (left) or region $\high$ (right) was used in the fit.}
        \label{figure:fit_LandH_BGL}
\end{figure}

\begin{figure}
\begin{center}
        \includegraphics[width=\linewidth]{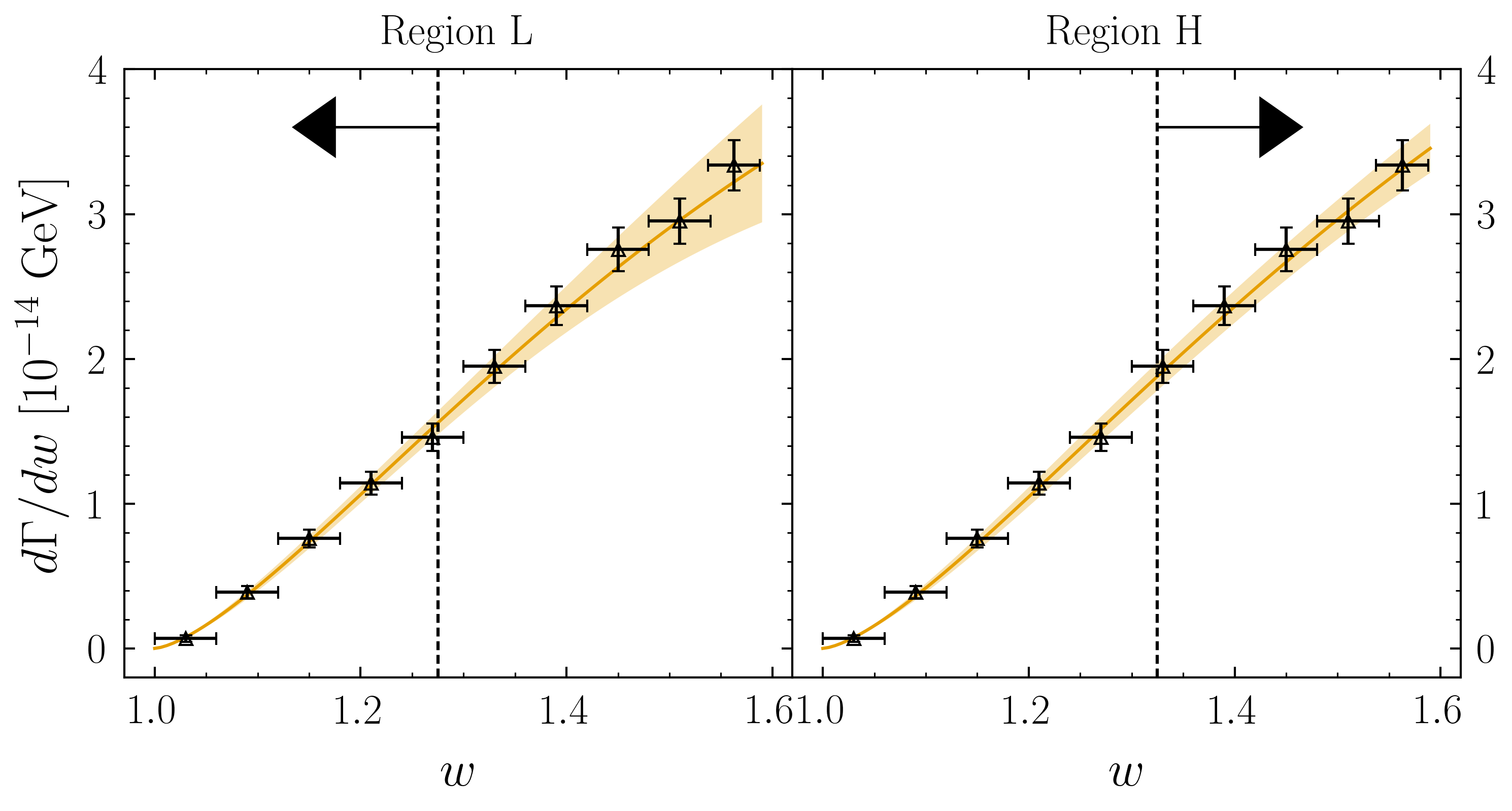}
\end{center}
        \caption{Same as Fig.~\ref{figure:fit_LandH_BGL} but for the BCL parameterization.}
        \label{figure:fit_LandH_BCL}
\end{figure}

The Belle data has $\nb=10$ bins, and we split the data in half between the $\low$ and $\high$ regions. We then fit the free parameters $a_{+,n}$ and $b_{+,n}$ to the $\low$ region data and the $\high$ region data separately. If we had perfect knowledge of $f_+$ in the $\low$ region we could reconstruct it in the $\high$ region and vice versa. An example of this is show for BGL with 2p in Fig.~\ref{figure:fit_LandH_BGL} and for BCL with 2p in Fig.~\ref{figure:fit_LandH_BCL}.

\begin{figure}[hbt!]
\begin{center}
         \includegraphics[width=\linewidth]{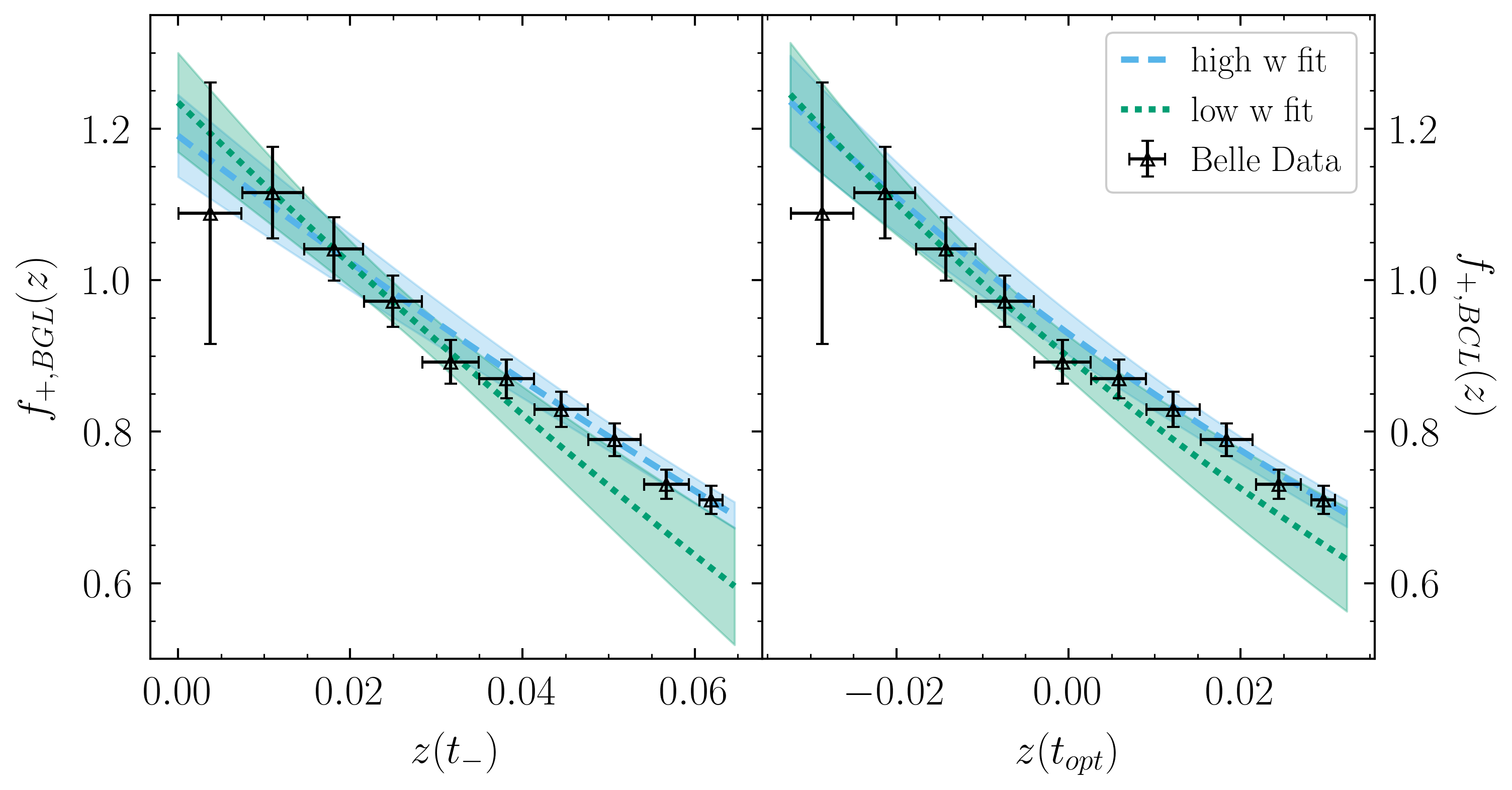}
\end{center}
        \caption{Plots of the form factor $f_+(z)$ vs $z$ using the results of our fits from regions $\low$ and $\high$. The results of our BGL fit with 2p (Left), the results of our BCL fit with 2p (Right). The red dashed line and red band indicate the fit from region $\high$, and the blue dashed line and blue band indicate the fit from region $\low$.}
        \label{figure:ff_2p}
\end{figure}

\begin{figure}
\begin{center}
        \includegraphics[width=\linewidth]{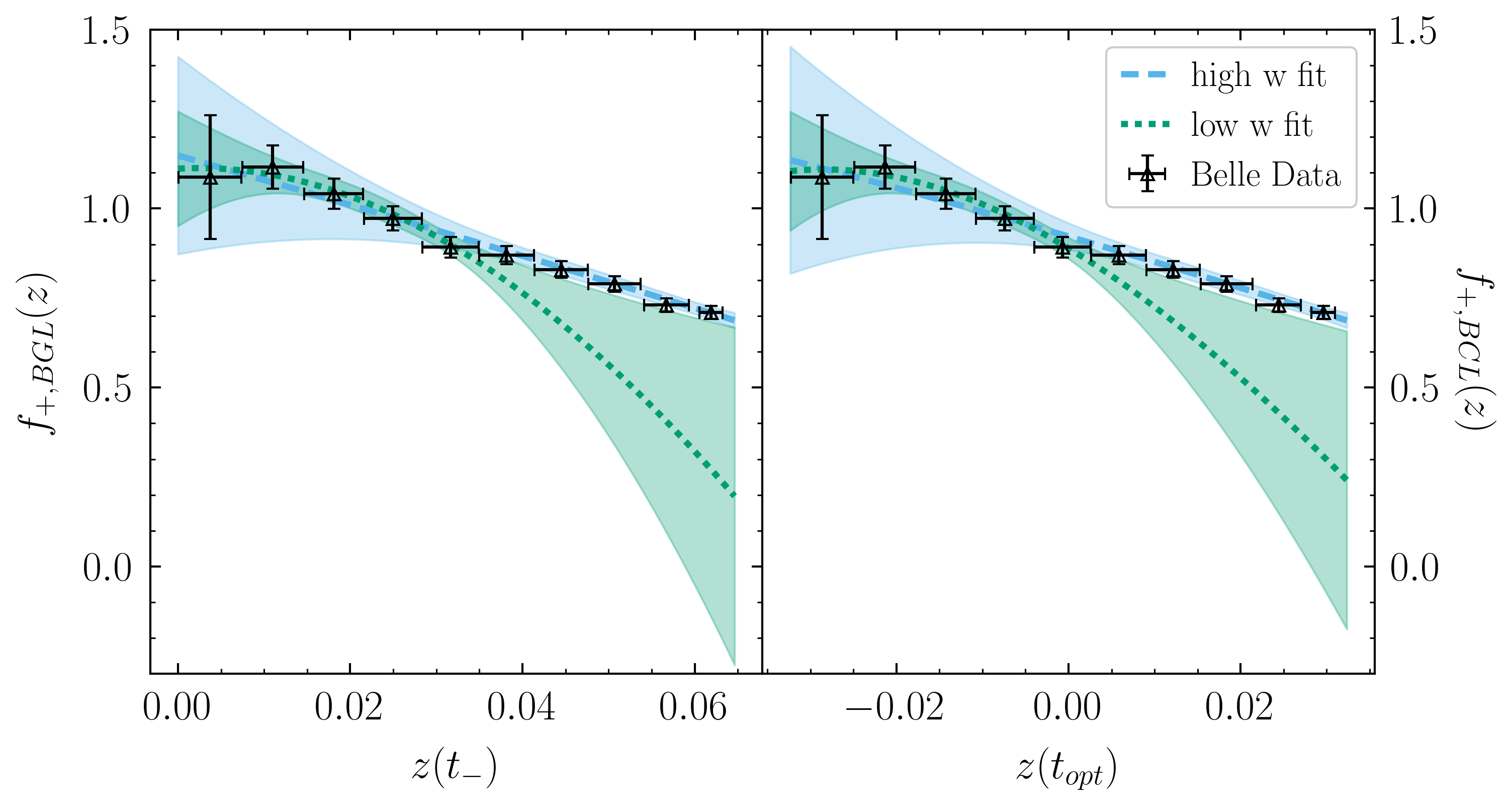}
\end{center}
        \caption{Same as Fig.~\ref{figure:ff_2p} but for 3p.}
        \label{figure:ff_3p}
\end{figure}

We carry out this method for BGL and BCL with 2p and 3p, and use the resulting parameters to plot the form factor $f_+(z)$ and convert the Belle data from differential decay width data to form factor data and include it in Figs.~\ref{figure:ff_2p} and~\ref{figure:ff_3p}. Both BGL and BCL have more overlap between the fits for 2p than for 3p. In the 3p fits case, the error band is very small in the $\high$ region and the error band increases in size as it moves into the $\low$ region, but the $\low$ fit has large error bands in both the $\low$ and $\high$ regions.

Now using the results of the fits for every parameterization with 1p, 2p and 3p, we are able to calculate the discrete and continuous metrics. The values of $C_0$ and $C_1$ that we calculate are listed in Table~\ref{tab:continuousmetric} for all parameterizations with 1p, 2p and 3p, and $D_1$ and $D_2$ are similarly shown in Table~\ref{tab:discretemetric}.

\begin{table}[hbt!]
\caption{\label{tab:continuousmetric}
The $C_0$ and $C_1$ values for BGL, BCL, NN and NT with 1p, 2p and 3p.}
\begin{ruledtabular}
\begin{tabular}{lll}
 & \multicolumn{1}{c}{$C_0$} & \multicolumn{1}{c}{$C_1$} \\
\colrule
BGL 1p & 15.76 & 39.44 \\ 
BGL 2p & 1.18 & 5.16 \\ 
BGL 3p & 17.56 & 91.76 \\ 
\hline
BCL 1p & 3.37 & 7.63 \\
BCL 2p & 0.63 & 2.72 \\
BCL 3p & 15.23 & 78.96 \\
\hline
NN 1p & 24.50 & 64.26 \\
NN 2p & 1.85 & 7.82 \\
NN 3p & 19.10 & 100.16 \\
\hline
NT 1p & 24.50 & 64.31 \\
NT 2p & 2.11 & 8.83 \\
NT 3p & 19.81 & 104.04 \\
\end{tabular}
\end{ruledtabular}
\end{table}

\begin{table}[hbt!]
\caption{\label{tab:discretemetric}
The $D_1$ and $D_2$ values for BGL, BCL, NN and NT with 1p, 2p and 3p.}
\begin{ruledtabular}
\begin{tabular}{lll}
 & \multicolumn{1}{c}{$D_1$} & \multicolumn{1}{c}{$D_2$} \\
\colrule
BGL 1p  & 41.85 & 10.95 \\
BGL 2p  & 5.78 & 4.76 \\
BGL 3p  & 104.64 & 117.68  \\
\hline
BCL 1p  & 8.02 & 1.90 \\
BCL 2p  & 3.00 & 1.74 \\
BCL 3p  & 89.85 & 100.37  \\
\hline
NN 1p  & 68.54 & 19.02 \\
NN 2p  & 8.88 & 8.21 \\
NN 3p  & 115.43 & 126.79  \\
\hline
NT 1p  & 68.54 & 19.02 \\
NT 2p  & 10.03 & 9.55 \\
NT 3p  & 119.70 & 131.76  \\
\end{tabular}
\end{ruledtabular}
\end{table}

Since $\Delta f(z)$ reflects the discrepancy between fits and extrapolations, the most self-consistent model is the one with the lowest values of $C_0$, $C_1$, $D_1$, and $D_2$. We see that the 2p values are all considerably lower than  the corresponding 1p or 3p values and are clearly preferred, which agrees with what we obtained with the AIC metrics. 
We define the mean values of the $C_0$, $C_1$, $D_1$, and $D_2$ for only the 2p models as $\bar{C}^{2p}_0 = 1.443$, $\bar{C}^{2p}_1 = 6.133$, $\bar{D}^{2p}_1 = 6.923$, and $\bar{D}^{2p}_2 = 6.065$. Comparing the $C_0$, $C_1$ and $D_1$ values from the 2p models with these mean values, the values from the BCL parameterization lie $55-56\%$ below the mean values, the values from the BGL parameterization lie $15-18\%$ below, the values from the NN parameterization lie $27-28\%$ above the mean values, and the values from the NT parameterization lie $43-46\%$ above the mean values. For the $D_2$ values from the 2p models, BCL lies $71\%$ below the mean value, BGL lies $21\%$ below the mean value, NN lies $35\%$ above the mean value and NT lies $57\%$ above the mean value. For each metric with 2p compared to the mean values with 2p, the BCL parameterization is significantly lower than the mean and appears to be the most self-consistent from the point of view of analyticity. 

We have also considered the effects of 
relaxing the threshold condition in BCL in \ref{appendixC} and changing the values of $t_0$, namely using 
$t_{opt}$ for BGL or $t_-$ for BCL in \ref{appendixB}. The tables make clear that these choices affect the metrics by at most a few percents and 
are essentially irrelevant.

\section{Analysis of self-consistency metrics}
\label{sec:analysis}

In this section, we discuss the relations among the new metrics. First we will compare different metrics for a given parameterization  and then we will compare the same metric but for different parameterizations. 

\subsection{Comparing the Metrics for a Given Parameterization}

The first comparison that we make is between $C_1$ and $C_0$. The ratios $C_1 / C_0$ can be found in Table~\ref{tab:C1overC0}. The values in the 2p row are consistent within $\sim 3\%$ of each other, and the values in the 3p row are within $\lesssim 1 \%$ of each other. In other words, the two metrics are proportional, with a proportionality constant which depends mostly on the number of parameters used in the fit. Note that the absolute normalization of 
$C_0$ or $C_1$ is not important. From the standard deviations in the binned data \cite{mainb2d}, we have 
\begin{equation}
\frac{1}{n_{bin}}\sum_{i=1}^{n_{bin}} \frac{1}{\sigma_i^2} \simeq
\frac{2.81}{(\frac{1}{n_{bin}}\sum_{i=1}^{n_{bin}} \sigma_i)^2},
\end{equation}
which partially explains that $C_1$ is larger than $C_0$.
We next look at $D_1 / C_1$. Since $D_1$ is a discrete version of $C_1$, we expect relative differences of the order of $1/n_{bin} =0.1$. We see that this is the case in Table~\ref{tab:D1overC1}.
\begin{table}[hbt!]
\caption{\label{tab:C1overC0}
The ratio of $C_1$ to $C_0$ calculated with BGL, BCL, NN and NT for 2p and 3p.}
\begin{ruledtabular}
\begin{tabular}{lllll}
 & \multicolumn{1}{c}{$C_{1,BGL}/C_{0,BGL}$} & \multicolumn{1}{c}{$C_{1,BCL}/C_{0,BCL}$} & \multicolumn{1}{c}{$C_{1,\textrm{NN}}/C_{0,\textrm{NN}}$} & \multicolumn{1}{c}{$C_{1,\textrm{NT}}/C_{0,\textrm{NT}}$} \\ 
\colrule
2p & 4.373 & 4.318 & 4.227 & 4.185 \\
3p & 5.226 & 5.185 & 5.244 & 5.252 \\
\end{tabular}
\end{ruledtabular}
\end{table}

\begin{table}[hbt!]
\caption{\label{tab:D1overC1}
The ratio of $D_1$ to $C_1$ calculated with BGL, BCL, NN and NT for 2p and 3p.}
\begin{ruledtabular}
\begin{tabular}{lllll}
 & \multicolumn{1}{c}{$D_{1,BGL}/C_{1,BGL}$} & \multicolumn{1}{c}{$D_{1,BCL}/C_{1,BCL}$} & \multicolumn{1}{c}{$D_{1,\textrm{NN}}/C_{1,\textrm{NN}}$} & \multicolumn{1}{c}{$D_{1,\textrm{NT}}/C_{1,\textrm{NT}}$} \\ 
\colrule
2p & 1.120 & 1.103 & 1.136 & 1.136 \\
3p & 1.140 & 1.138 & 1.154 & 1.151 \\ 
\end{tabular}
\end{ruledtabular}
\end{table}


In summary, we found that the three metrics $C_0$, 
$C_1$ and $D_1$ provide consistent estimates of the departure from analyticity. 
For instance, we could just consider $D_1$ which is easier to calculate from experimental binned data. So far, we have ignored correlations among the bins. Table II of Ref. \cite{mainb2d} shows that these correlations are significant which motivated the introduction of $D_2$. 
The ratios $D_2 / D_1$ are provided in Table~\ref{tab:D2overD1}. We see that for 2p 
the ratios have a stronger dependence on the 
parameterization which amplifies the discrimination (a lower $D_1$ means an even lower $D_2$). On the other hand for the suboptimal choice 3p, the ratio is about 1.1 in the four cases and $D_2$ does not provide new information.

\begin{table}[hbt!]
\caption{\label{tab:D2overD1}
The ratio of $D_2$ to $D_1$ calculated with BGL, BCL, NN and NT for 2p and 3p.}
\begin{ruledtabular}
\begin{tabular}{lllll}
 & \multicolumn{1}{c}{$D_{2,BGL}/D_{1,BGL}$} & \multicolumn{1}{c}{$D_{2,BCL}/D_{1,BCL}$} & \multicolumn{1}{c}{$D_{2,\textrm{NN}}/D_{1,\textrm{NN}}$} & \multicolumn{1}{c}{$D_{2,\textrm{NT}}/D_{1,\textrm{NT}}$} \\ 
\colrule
2p & 0.824 & 0.580 & 0.925 & 0.952 \\ 
3p & 1.125 & 1.117 & 1.098 & 1.101 \\ 
\end{tabular}
\end{ruledtabular}
\end{table}

The values in the 3p row are consistent within $\lesssim 1 \%$ of the other values in the 3p row, however we see that the $D_2 / D_1$ ratio for 2p is smaller for BCL than it is for BGL, NN or NT by roughly $43\%$. It is great that these comparisons show consistency, and they show that the $D_2$ metric provides the most information to discriminate between the parameterizations.

\subsection{Comparing the Same Metric for Different Parameterizations}

Since the NN and NT parameterizations are not ever mentioned in the literature, at this point we ignore them and focus again on BGL and BCL as they were shown to be preferred over NN or NT by every metric we considered. Now comparing the same metric between the BGL and BCL parameterizations, the ratios $C_{i,BCL} / C_{i,BGL}$ for $i=0,1$ are given in Table~\ref{tab:CioverCi} and $D_{j,BCL} / D_{j,BGL}$ for $j=1,2$ in Table~\ref{tab:DioverDi}. For this, we recalculated the BGL parameterization with $t_0 = t_{opt}$ and the BCL parameterization with $t_0 = t_-$, in order to have the $z$-expansion consistent when comparing different parameterizations.

\begin{table}[hbt!]
\caption{\label{tab:CioverCi}
The ratio of $C_{i,BCL}$ to $C_{i,BGL}$ calculated for both 2p and 3p with both choices of $t_0$.}
\begin{ruledtabular}
\begin{tabular}{lllll}
 & \multicolumn{2}{c}{$C_{0,BCL}/C_{0,BGL}$} & \multicolumn{2}{c}{$C_{1,BCL}/C_{1,BGL}$} \\ 
 & $t_0 = t_{-}$ & $t_0 = t_{opt}$ & $t_0 = t_{-}$ & $t_0 = t_{opt}$ \\
\colrule
2p & 0.536 & 0.545 & 0.523 & 0.532 \\ 
3p & 0.862 & 0.866 & 0.855 & 0.860 \\
\end{tabular}
\end{ruledtabular}
\end{table}

\begin{table}[hbt!]
\caption{\label{tab:DioverDi}
The ratio of $D_{j,BCL}$ to $D_{j,BGL}$ calculated for both 2p and 3p with both choices of $t_0$.}
\begin{ruledtabular}
\begin{tabular}{lllll}
 & \multicolumn{2}{c}{$D_{1,BCL}/D_{1,BGL}$} & \multicolumn{2}{c}{$D_{2,BCL}/D_{2,BGL}$} \\ 
 & $t_0 = t_{-}$ & $t_0 = t_{opt}$ & $t_0 = t_{-}$ & $t_0 = t_{opt}$ \\
\colrule
2p & 0.515 & 0.524 & 0.360 & 0.370 \\
3p & 0.854 & 0.859 & 0.849 & 0.854 \\
\end{tabular}
\end{ruledtabular}
\end{table}

It is important is observe that the values in Tables~\ref{tab:CioverCi} and~\ref{tab:DioverDi} for the 3p rows are all consistent within $\lesssim 1 \%$ of the other values in the 3p rows. It is also important to see that the values in the 2p rows are very similar for $C_{0,BCL}/C_{0,BGL}$, $C_{1,BCL}/C_{1,BGL}$ and $D_{1,BCL}/D_{1,BGL}$, however there is decrease of roughly $36\%$ in the 2p values for $D_{2,BCL}/D_{2,BGL}$. This shows the consistency of our defined metrics, and shows that $C_0$, $C_1$ and $D_1$ offer a similar amount of information compared to $D_2$, which possibly contains more information about the fits because it is the only metric to differ in these categories when compared.

\section{Conclusion}

In conclusions, we investigated the BGL and BCL parameterizations of the form factor used in the differential decay rate of $B \rightarrow D \ell \nu_{\ell}$. 
With the experimental binned data collected by the Belle collaboration~\cite{mainb2d}, 
we found that the standard $\chi^2$ and $\chi^2_{\nu}$ do not provide us with enough information to distinguish between BGL vs BCL or 2p vs 3p. The AIC clearly favors 2p over 1p or 3p but the differences between BGL and BCL are too small to be meaningful.

We introduced four metrics or ``cost functions" ($C_0,\ C_1,\ D_1$ and $D_2$) that measure the 
discrepancy between fits and extrapolations of the regular parts of form factors in 
the high and low parts of the kinematic range. Given the analyticity of these regular parts, 
a perfect fit in one region would provide a unique and perfect analytical continuation in the other region and vice-versa. The first metric ($C_0$) is a dimensionless $L^2$ norm of the discrepancy. $C_1$ is a locally weighted version of $C_0$ that favors the kinematic regions with smaller experimental uncertainties. $D_1$ is a discretized version of $C_1$ which can be implemented directly from the experimental binned data. $D_2$ 
is an extension of $D_1$ that incorporate the 
correlations among the bins. In view of the significant bin correlations \cite{mainb2d}, 
$D_2$ should be a better measure than $D_1$.
$C_0,\ C_1,$ and $D_1$ provide very similar and consistent discriminations while $D_2$ somehow amplifies the discriminations for 2p.

All the metrics strongly favor 2p over 3p. 
A possible interpretation is that the experimental uncertainties prevent an accurate determination of the quadratic corrections and that one partially extrapolates the experimental noise which is not an analytical function of $z$. 
All the metrics favor 2p over 1p. 
Except for BCL, the metrics are about twice larger for 1p. 
It could be that the corrections to the constant approximation 
are significant and result in significantly different constant approximations in the high and low $z$ regions. On the other hand, for BCL it appears that 1p is a better approximation than for the other parameterization. 

Focusing on the 2p results, we find a finer resolution among parameterizations. For all the metrics, we observe smaller values for BCL
than for the other parameterizations.
In addition BGL does better than no 
prefactor. 
It is possible that in the case considered here, the BCL prefactor captures the features of the actual form factor in a slightly better way. This is hinted by the 
fact that a constant approximation has a significantly smaller $\chi^2$ 
for BCL. This observation may be anecdotal and study of other cases
should bring more light on the question. 
We also found that other choices such as the the value of $t_0$ or the imposition of a threshold condition have a marginal impact on the values of the metrics. 

It should be emphasized that all the  metrics measure discrepancies among fits and not closeness to data. 
It might be possible to include them
in augmented $\chi^2$ \cite{AICdefinition_2020,AICdefinition_2022}, however, determining the coefficient in front of the metric is a nontrivial task. It should also be noted that very recently, Bayesian inference methods have been used to deal with the truncation question
\cite{flynn2023bayesian} and applied to $B_s\rightarrow K\ell \nu_{\ell}$ \cite{flynn2023exclusive}. These methods consider higher order expansions and provide results in agreement with other calculations based on unitarity  \cite{Di_Carlo_2021}. 
It would be very interesting to repeat our analysis using this Bayesian inference procedure for the two sets of bins considered here separately and compare alternative higher-order expansions with our metrics.

So far our calculations of the metrics have been limited to one set of experimental data \cite{mainb2d} for $B\rightarrow D \ell \nu_{\ell}$ 
and it is premature to draw general conclusions. 
Applying the method to other process involving the the z-expansion should 
help identifying more general properties. 
The $z$-expansion has also been used extensively in the study of nucleon form factors. Various
neutrino-deuteron scattering experiments have been combined to extract the $z-$expansion of the
isovector axial nucleon form factor from experiment \cite{Meyer:2016oeg}. The $z$-expansion has also been used to parameterize lattice calculations of the same quantity, see for instance  \cite{Ishikawa:2018rew,Kronfeld:2019nfb,Jang:2019vkm,Jang:2019jkn,RQCD:2019jai,Alexandrou:2020okk,Park:2021ypf,Ruso:2022qes,Djukanovic:2022wru} 
and more references in a recent review article \cite{Meyer:2022mix}.
These parameterizations have been used to incorporate 
nucleon effects in the calculations of neutrino-nucleus cross section \cite{Simons:2022ltq}. 
The method that we proposed can be applied to nucleon form factors as long as one can perform 
new fits in distinct kinematic regions. This is feasible for binned data, but if extrapolations procedures are involved, such as the continuum limit in lattice calculations, all the details of the 
existing procedure need to be repeated in kinematic subregions.

\begin{acknowledgments}
This
research was supported in part by Department of Energy under Award Numbers DOE grant
DE-SC0010113. We thank R. Van de Water for emphasizing the need of a metric involving covariances and for comments on the presentation.
We thank M. Wagman for comments on the AIC criterion and for comments on the manuscript. We thank A. Kronfeld, and  F. Herren for valuable discussions and A. Juttner and O. Witzel for pointing out recent references.  
\end{acknowledgments}

\appendix
\section{Our Calculated Fit Parameters}\label{appendixA}

For completeness, we list the fit parameters that were the result of our fits to the Belle data. For 1p, 2p and 3p, we show the BGL and BCL fit parameters in~\ref{tab:global_fit_params} and the NN and NT parameters in~\ref{tab:global_fit_params_NN_NT}. We also provide the ratios of the fit parameters for BGL and BCL in~\ref{tab:global_fit_params_ratios} and for NN and NT in~\ref{tab:global_fit_params_ratios_NN_NT}.

\begin{table}[hbt!]
\caption{\label{tab:global_fit_params}
The $a_{+,n}$ and $b_{+,n}$ values that came from the global fit of BGL and BCL.}
\begin{ruledtabular}
\begin{tabular}{lllllll}
 & \multicolumn{3}{c}{BGL} & \multicolumn{3}{c}{BCL} \\
  & {$a_{+,0}$}&{$a_{+,1}$}&{$a_{+,2}$}&{$b_{+,0}$}&{$b_{+,1}$}&{$b_{+,2}$} \\
\colrule
1p   & 0.00804(19) & --- & --- & 0.703(16) & --- & ---  \\
2p   & 0.01238(41) & -0.0654(58) & --- & 0.773(19) & -2.41(42) & ---  \\
3p   & 0.01248(67)  & -0.071(31)  & 0.07(37) & 0.775(20) & -2.28(64) & -8(28) \\
\end{tabular}
\end{ruledtabular}
\end{table}

\begin{table}[hbt!]
\caption{\label{tab:global_fit_params_ratios} The ratios $a_{+,n+1}/a_{+,n}$ and $b_{+,n+1},b_{+,n}$ using the central values of the parameters from~\ref{tab:global_fit_params}.}
\begin{ruledtabular}
\begin{tabular}{lllll}
 & \multicolumn{2}{c}{BGL} & \multicolumn{2}{c}{BCL} \\
  &{$a_{+,1} / a_{+,0}$}&{$a_{+,2} / a_{+,1}$}&{$b_{+,1} / b_{+,0}$}&{$b_{+,2} / b_{+,1}$}\\
\colrule
2p   & -5.281 & --- & -3.112 & ---  \\
3p   & -5.710  & -0.99998 & -2.942 & 3.716 \\
\end{tabular}
\end{ruledtabular}
\end{table}

\begin{table}[hbt!]
\caption{\label{tab:global_fit_params_NN_NT}
The $a_{+,n}$ and $b_{+,n}$ values that came from the global fit of NN and NT.}
\begin{ruledtabular}
\begin{tabular}{lllllll}
 & \multicolumn{3}{c}{NN} & \multicolumn{3}{c}{NT} \\
  & {$a_{+,0}$}&{$a_{+,1}$}&{$a_{+,2}$}&{$b_{+,0}$}&{$b_{+,1}$}&{$b_{+,2}$} \\
\colrule
1p   & 0.666(17) & --- & --- & 0.666(17) & --- & ---  \\
2p   & 1.154(37) & -7.22(53) & --- & 0.921(23) & -7.14(53) & ---  \\
3p   & 1.181(61)  & -8.7(2.8)  & 18(33) & 0.917(24) & -7.56(82) & 18(32) \\
\end{tabular}
\end{ruledtabular}
\end{table}

\begin{table}[hbt!]
\caption{\label{tab:global_fit_params_ratios_NN_NT} The ratios $a_{+,n+1}/a_{+,n}$ and $b_{+,n+1},b_{+,n}$ using the central values of the parameters from~\ref{tab:global_fit_params_NN_NT}.}
\begin{ruledtabular}
\begin{tabular}{lllll}
 & \multicolumn{2}{c}{NN} & \multicolumn{2}{c}{NT} \\
  &{$a_{+,1} / a_{+,0}$}&{$a_{+,2} / a_{+,1}$}&{$b_{+,1} / b_{+,0}$}&{$b_{+,2} / b_{+,1}$}\\
\colrule
2p   & -6.257 & --- & -7.750 & ---  \\
3p   & -7.404  & -2.088 & -8.236 & -2.376 \\
\end{tabular}
\end{ruledtabular}
\end{table}

\section{Investigating the \texorpdfstring{$t_0$}{} Parameter}\label{appendixB}

The BGL and BCL parameterizations use different choices for $t_0$, although the value of $t_0$ does not affect the size of the $z$ range but it does affect the center of the $z$ range. We investigated the effect of $t_0$ on our metrics by calculating all the metrics using both choices of $t_0$. The $\chi^2$, $\chi^2_{\nu}$ and AIC values can be found in Table~\ref{tab:all_metrics1}, and the $C_0$, $C_1$, $D_1$ and $D_2$ values can be found in Table~\ref{tab:all_metrics2}.

\begin{table}[hbt!]
\caption{\label{tab:all_metrics1}
Same as Tables~\ref{tab:DiffDec_chi2} and~\ref{tab:DiffDec_AIC}, but every value is calculated with both choices of $t_0$.}
\begin{ruledtabular}
\begin{tabular}{lllllll}
 & \multicolumn{2}{c}{$\chi^2$} & \multicolumn{2}{c}{$\chi^2_{\nu}$} & \multicolumn{2}{c}{AIC} \\ 
 & $t_0 = t_{-}$ & $t_0 = t_{opt}$ & $t_0 = t_{-}$ & $t_0 = t_{opt}$ & $t_0 = t_{-}$ & $t_0 = t_{opt}$ \\
\colrule
BGL 1p & 99 & 99 & 11 & 11 & 101 & 101 \\ 
BGL 2p & 4.56 & 4.56 & 0.57 & 0.57 & 8.56 & 8.56 \\ 
BGL 3p & 4.55 & 4.55 & 0.65 & 0.65 & 10.55 & 10.55 \\
\hline
BCL 1p & 33.3 & 33.3 & 3.7 & 3.7 & 35.3 & 35.3 \\
BCL 2p & 4.64 & 4.64 & 0.58 & 0.58 & 8.64 & 8.64 \\
BCL 3p & 4.55 & 4.55 & 0.65 & 0.65 & 10.55 & 10.55 \\ 
\hline
NN 1p & 135 & 135 & 15 & 15 & 137 & 137 \\
NN 2p & 4.88 & 4.88 & 0.61 & 0.61 & 8.88 & 8.88 \\
NN 3p & 4.55 & 4.55 & 0.65 & 0.65 & 10.55 & 10.55 \\ 
\hline
NT 1p & 135 & 135 & 15 & 15 & 137 & 137 \\
NT 2p & 4.96 & 5.04 & 0.62 & 0.63 & 8.96 & 9.04 \\
NT 3p & 4.55 & 4.55 & 0.65 & 0.65 & 10.55 & 10.55 \\ 
\end{tabular}
\end{ruledtabular}
\end{table}

\begin{table}[hbt!]
\caption{\label{tab:all_metrics2}
Same as Tables~\ref{tab:continuousmetric} and~\ref{tab:discretemetric} but every value is calculated with both choices of $t_0$.}
\begin{ruledtabular}
\begin{tabular}{lllllllll}
 & \multicolumn{2}{c}{$C_0$} & \multicolumn{2}{c}{$C_1$} & \multicolumn{2}{c}{$D_1$} & \multicolumn{2}{c}{$D_2$} \\ 
 & $t_0 = t_{-}$ & $t_0 = t_{opt}$ & $t_0 = t_{-}$ & $t_0 = t_{opt}$ & $t_0 = t_{-}$ & $t_0 = t_{opt}$ & $t_0 = t_{-}$ & $t_0 = t_{opt}$ \\
\colrule
BGL 1p & 15.76 & 15.38 & 39.44 & 38.44 & 41.85 & 41.76 & 10.95 & 10.64 \\
BGL 2p & 1.18 & 1.16 & 5.16 & 5.11 & 5.78 & 5.72 & 4.76 & 4.69 \\
BGL 3p & 17.56 & 17.58 & 91.76 & 91.83 & 104.64 & 104.54 & 117.68 & 117.57 \\ 
\hline
BCL 1p & 3.37 & 3.37 & 7.62 & 7.63 & 8.02 & 8.02 & 1.90 & 1.90 \\
BCL 2p & 0.63 & 0.63 & 2.70 & 2.72 & 2.98 & 3.00 & 1.71 & 1.74 \\
BCL 3p & 15.13 & 15.23 & 78.44 & 78.96 & 89.39 & 89.85 & 99.85 & 100.37 \\ 
\hline
NN 1p & 24.50 & 24.50 & 64.26 & 64.26 & 68.54 & 68.54 & 19.02 & 19.02 \\
NN 2p & 1.85 & 1.87 & 7.82 & 7.88 & 8.88 & 8.95 & 8.21 & 8.29 \\
NN 3p & 19.10 & 19.19 & 100.16 & 100.66 & 115.43 & 115.81 & 126.79 & 127.23 \\ 
\hline
NT 1p & 24.50 & 24.50 & 64.31 & 64.31 & 68.54 & 68.54 & 19.02 & 19.02 \\
NT 2p & 2.09 & 2.11 & 8.72 & 8.83 & 9.92 & 10.03 & 9.42 & 9.55 \\
NT 3p & 19.69 & 19.81 & 103.42 & 104.04 & 119.70 & 119.70 & 131.16 & 131.76 \\ 
\end{tabular}
\end{ruledtabular}
\end{table}

We find that the choice of $t_0$ has negligible effects on the $\chi^2$, $\chi^2_{\nu}$ and AIC metrics at all with the precision that we consider. However, the $C_0$ and $C_1$ as well as the $D_1$ and $D_2$ metrics have some minor differences based on the choice of $t_0$, but the differences are on the order of $1\%$. This confirms that the main role of $t_0$ is to set the central value of the $z$ range.

\section{BCL With No Threshold Condition}\label{appendixC}

The threshold condition from~\cite{originalBCL} comes from $z(t_+, t_0) = -1$ which can be seen in Eq.~\eqref{eq:zexp_equation}, and from the fact that $(z+1) \sim \textrm{const.} \times (q^2 - t_+)^{1/2}$ near $z=-1$. Then the threshold condition is,

\begin{equation}
    \biggr[ \frac{d f_+}{d z} \biggr]_{z=-1} = 0 .
\label{eq:threshold_condition}
\end{equation}

\noindent
We investigate the effect of the threshold condition by reproducing our results using the BCL parameterization with no threshold condition, which we call BCL$^*$ has the form,

\begin{equation}
    f_{+,BCL^*}(z) = \frac{1}{1 - q^2(z)/m_{B_c^*}^2} \sum^{N}_{n=0} b_{+,n} z^n . \\
\end{equation}

\noindent
Using this BCL$^*$, we recalculate all the values in Tables~\ref{tab:all_metrics1} and~\ref{tab:all_metrics2} and display the results below. 

\begin{table}[hbt!]
\caption{\label{tab:BCLstar_metrics1}
Same as Table~\ref{tab:all_metrics1} but using BCL$^*$.}
\begin{ruledtabular}
\begin{tabular}{lllllll}
 & \multicolumn{2}{c}{$\chi^2$} & \multicolumn{2}{c}{$\chi^2_{\nu}$} & \multicolumn{2}{c}{AIC} \\ 
 & $t_0 = t_{-}$ & $t_0 = t_{opt}$ & $t_0 = t_{-}$ & $t_0 = t_{opt}$ & $t_0 = t_{-}$ & $t_0 = t_{opt}$ \\
\colrule
BCL$^*$ 1p & 33.3 & 33.3 & 3.7 & 3.7 & 35.3 & 35.3 \\
BCL$^*$ 2p & 4.64 & 4.64 & 0.58 & 0.58 & 8.64 & 8.64 \\
BCL$^*$ 3p & 4.55 & 4.55 & 0.65 & 0.65 & 10.55 & 10.55 \\ 
\end{tabular}
\end{ruledtabular}
\end{table}

\begin{table}[hbt!]
\caption{\label{tab:BCLstar_metrics2}
Same as Table~\ref{tab:all_metrics2} but using BCL$^*$.}
\begin{ruledtabular}
\begin{tabular}{lllllllll}
 & \multicolumn{2}{c}{$C_0$} & \multicolumn{2}{c}{$C_1$} & \multicolumn{2}{c}{$D_1$} & \multicolumn{2}{c}{$D_2$} \\ 
 & $t_0 = t_{-}$ & $t_0 = t_{opt}$ & $t_0 = t_{-}$ & $t_0 = t_{opt}$ & $t_0 = t_{-}$ & $t_0 = t_{opt}$ & $t_0 = t_{-}$ & $t_0 = t_{opt}$ \\
\colrule
BCL$^*$ 1p & 3.37 & 3.37 & 7.62 & 7.63 & 8.04 & 8.04 & 1.93 & 1.93 \\
BCL$^*$ 2p & 0.63 & 0.61 & 2.65 & 2.55 & 2.96 & 2.84 & 1.61 & 1.54 \\
BCL$^*$ 3p & 17.03 & 17.88 & 88.14 & 77.00 & 102.68 & 89.58 & 109.22 & 96.28 \\ 
\end{tabular}
\end{ruledtabular}
\end{table}

\noindent
These values are mostly identical to the BCL values shown in Tables~\ref{tab:all_metrics1} and~\ref{tab:all_metrics2}, with the only differences appearing in $C_0$, $C_1$, $D_1$ and $D_2$.


%

\end{document}